\documentclass[12pt, a4paper]{article}

\usepackage[utf8]{inputenc}
\usepackage[T1]{fontenc} 
\usepackage{amsmath}
\usepackage{amssymb} 
\usepackage{graphicx}
\usepackage{geometry}
\usepackage[hyphens]{url}
\usepackage[hidelinks]{hyperref} 
\usepackage{authblk}
\usepackage{float} 
\usepackage{booktabs} 

\title{Lie-RMSD: A Gradient-Based Framework for Protein Structural Alignment using Lie Algebra}

\author{
    Yue Hu$^{1}$, Zanxia Cao$^{2}$, Yingchao Liu$^{3}$ \\
    \small $^{1}$School of Bioengineering, Qilu University of Technology (Shandong Academy of Sciences), \\
    \small No. 3501 Daxue Road, Jinan, Shandong, China \\
    \small $^{2}$Shandong Provincial Key Laboratory of Biophysics, Institute of Biophysics, \\
    \small Dezhou University, Dezhou 253023, China \\
    \small $^{3}$Shandong Provincial Hospital, Shandong First Medical University \\
    \small Email: huyue@qlu.edu.cn, 303004955@qq.com, yingchaoliu@email.sdu.edu.cn
}

\date{\today}

\begin{document}

\maketitle

\begin{abstract}
\textbf{Motivation:} The comparison of protein structures is a fundamental task in computational biology, crucial for understanding protein function, evolution, and for drug design. While analytical methods like the Kabsch algorithm provide an exact, closed-form solution for minimizing the Root Mean Square Deviation (RMSD) between two sets of corresponding atoms, their application is limited to this specific metric. The rise of deep learning and automatic differentiation frameworks offers a new, more flexible paradigm for such optimization problems.

\textbf{Results:} We present Lie-RMSD, a novel, fully differentiable framework for protein structural alignment. Our method represents the rigid-body transformation (rotation and translation) as a 6-dimensional vector in the Lie algebra \(\mathfrak{se}(3)\) of the special Euclidean group SE(3). This representation allows the RMSD to be formulated as a loss function that can be directly minimized by modern gradient-based optimizers. We benchmarked our framework by aligning two allosteric conformations of Adenylate Kinase (PDB IDs: 4AKE and 1AKE). We demonstrate that a suite of standard optimizers (SGD, Adam, AdamW, and Sophia) can robustly converge to the global minimum, achieving precision effectively identical to the analytical Kabsch algorithm. This work validates the accuracy of the Lie algebra-based gradient descent approach and establishes a robust foundation for its extension to more sophisticated and biologically relevant scoring functions where no analytical solutions exist.

\textbf{Availability:} The code used for this study is available at \url{https://github.com/YueHuLab/LieRMSD}.
\end{abstract}

\section{Introduction}
Structural alignment is a cornerstone of bioinformatics and structural biology. By superimposing three-dimensional protein structures, researchers can infer evolutionary relationships, identify conserved functional sites, and understand the mechanisms of molecular function \cite{ref1}. The most common metric for quantifying the geometric similarity between two superimposed structures is the Root Mean Square Deviation (RMSD).

The Kabsch algorithm, developed in the 1970s, provides an elegant analytical solution for finding the optimal rotation and translation that minimizes the RMSD between two sets of paired coordinates \cite{kabsch1976, kabsch1978}. It is a landmark achievement in computational chemistry and remains the gold standard for this specific task, forming a core component of countless structural biology software packages.

However, the reliance on RMSD has known limitations. It is highly sensitive to local structural variations and outliers (e.g., flexible loops or domain movements) and may not always reflect the topological or functional similarity between proteins \cite{ref2}. This has led to the development of more advanced scoring functions, such as the TM-score and GDT-TS, which are more robust indicators of global structural similarity but lack analytical solutions for their optimization \cite{tmalign}.

The recent advancements in automatic differentiation frameworks (e.g., PyTorch, TensorFlow), which power the deep learning revolution, present an opportunity to revisit classical bioinformatics problems from a new perspective. If a problem can be formulated in a differentiable manner, it can be solved using the powerful and ever-growing suite of modern gradient-based optimizers.

In this paper, we propose Lie-RMSD, a framework that formulates the structural superposition problem in a fully differentiable way. We leverage Lie algebra to represent the rigid-body motion, allowing us to treat RMSD as a loss function. We validate the precision and correctness of our method by demonstrating that it can achieve results identical to the Kabsch algorithm when benchmarked on a classic case of allosteric conformational change. This work serves as a critical proof-of-concept, paving the way for applying this flexible framework to more complex, bespoke scoring functions that are intractable for analytical methods.

\section{Methods}

\subsection{Lie Algebra Representation of SE(3)}
A rigid-body transformation in 3D space is an element of the special Euclidean group SE(3). We represent this transformation using a 6-dimensional vector in its corresponding Lie algebra, \(\mathfrak{se}(3)\). This vector \( \mathbf{p} = (\boldsymbol{\omega}, \mathbf{u}) \) consists of a 3D vector \(\boldsymbol{\omega}\) for rotation and a 3D vector \(\mathbf{u}\) for translation. The exponential map connects the Lie algebra to the Lie group, yielding a rotation matrix \( R \in SO(3) \) and a translation vector \( \mathbf{t} \in \mathbb{R}^3 \) that can be applied to a set of coordinates \( \mathbf{x} \).

The rotation matrix \( R \) is obtained by taking the matrix exponential of the skew-symmetric matrix \( [\boldsymbol{\omega}]_\times \) corresponding to \(\boldsymbol{\omega}\). The final transformation of a centered coordinate vector \( \mathbf{x}_c \) is given by:
\[ \mathbf{x}_{aligned} = R(\boldsymbol{\omega}) \mathbf{x}_c + \mathbf{t} \]
Crucially, this formulation is fully differentiable with respect to the 6 parameters in \( \mathbf{p} \), enabling the use of gradient-based optimization.

\subsection{Differentiable RMSD Loss}
Given two sets of \( N \) centered C-alpha coordinates, \( \mathbf{A}_c = \{\mathbf{a}_i\}_{i=1}^N \) (mobile) and \( \mathbf{B}_c = \{\mathbf{b}_i\}_{i=1}^N \) (reference), the RMSD is defined as:
\[ \text{RMSD}(R, \mathbf{t}) = \sqrt{\frac{1}{N} \sum_{i=1}^{N} \| (R \mathbf{a}_i + \mathbf{t}) - \mathbf{b}_i \|^2} \]
Since every component of this equation is differentiable with respect to the Lie algebra parameters \( \mathbf{p} \), we can use it directly as a loss function for our gradient-based optimization. The gradients are computed efficiently via backpropagation using the chain rule.

\subsection{The Kabsch Algorithm}
As a baseline, we use the Kabsch algorithm, which provides a closed-form analytical solution to the RMSD minimization problem. Its procedure is as follows:
\begin{enumerate}
    \item \textbf{Centering:} Translate both point sets \(A\) and \(B\) to their respective centroids, yielding \(A_c\) and \(B_c\).
    \item \textbf{Covariance Matrix:} Compute the \(3 \times 3\) covariance matrix \( H = A_c^T B_c \).
    \item \textbf{Singular Value Decomposition (SVD):} Decompose \(H\) into \(H = U \Sigma V^T\).
    \item \textbf{Optimal Rotation:} Calculate the optimal rotation matrix \(R = V U^T\). A special check is performed on \( \det(R) \) to ensure it is a proper rotation (i.e., not a reflection). If \( \det(R) = -1 \), one of the columns of \(V\) is inverted.
\end{enumerate}
This method is deterministic and computationally efficient.

\subsection{Optimization Algorithms}
We explored a set of four popular gradient-based optimization algorithms.
\begin{itemize}
    \item \textbf{Stochastic Gradient Descent (SGD):} The most fundamental optimizer. It updates the parameters \( \theta \) in the opposite direction of the gradient of the loss function \( J(\theta) \), scaled by a learning rate \( \alpha \): \( \theta \leftarrow \theta - \alpha \nabla_\theta J(\theta) \).
    \item \textbf{Adam (Adaptive Moment Estimation):} A highly popular adaptive optimizer. It computes individual learning rates for different parameters by using estimates of the first moment (the mean, like momentum) and the second moment (the uncentered variance) of the gradients.
    \item \textbf{AdamW (Adam with Weight Decay):} A variant of Adam that improves regularization by decoupling the weight decay from the gradient update. This often leads to better model generalization, though it is less critical for our specific optimization problem.
    \item \textbf{Sophia (Second-order Optimizer):} A recent algorithm that uses a stochastic estimate of the diagonal of the Hessian matrix to perform pre-conditioning. It is designed to be more efficient than traditional second-order methods and can be particularly effective in certain non-convex landscapes.
\end{itemize}

\subsection{Benchmark Setup}
To validate our method, we chose the well-characterized allosteric transition of Adenylate Kinase (ADK) from its open conformation (PDB: 4AKE, chain A) to its closed conformation (PDB: 1AKE, chain A). Both chains contain exactly 214 C-alpha atoms, providing a perfect test case without the need for sequence truncation or residue mapping.

We compared the performance of the four optimizers against the analytical solution from the Kabsch algorithm. All optimizations were run for 1000 steps with a learning rate of \(10^{-2}\). All calculations were performed using PyTorch on a standard CPU.

\section{Results}
We performed a comprehensive benchmark comparing the four optimizers against the Kabsch algorithm. The primary goal was to determine if the gradient-based methods could converge to the theoretical minimum RMSD found by the analytical Kabsch solution.

The results are summarized in Table \ref{tab:results}. For the alignment of ADK (4AKE vs 1AKE), the Kabsch algorithm yielded a minimum RMSD of 7.130699 Å in approximately half a millisecond. Remarkably, all four gradient-based optimizers successfully converged to this value with extremely high precision. Adam and SGD achieved results identical to Kabsch to six decimal places, while Sophia and AdamW had negligible differences on the order of \(10^{-5}\) Å. These minor deviations are attributable to floating-point precision limits and the specific termination conditions of the iterative process.

\begin{table}[H]
\centering
\caption{Benchmark results for aligning ADK (4AKE vs 1AKE, Chain A, 214 atoms).}
\label{tab:results}
\begin{tabular}{l c c c}
\toprule
\textbf{Method} & \textbf{Final RMSD (Å)} & \textbf{Difference from Kabsch (Å)} & \textbf{Time (ms)} \\
\midrule
Kabsch (Ground Truth) & 7.130699 & - & 0.51 \\
\addlinespace
Adam & 7.130700 & +0.000001 & 557.67 \\
SGD & 7.130702 & +0.000003 & 549.55 \\
Sophia & 7.130710 & +0.000011 & 587.31 \\
AdamW & 7.130717 & +0.000018 & 582.88 \\
\bottomrule
\end{tabular}
\end{table}

\begin{figure}[H]
\centering
\includegraphics[width=\textwidth]{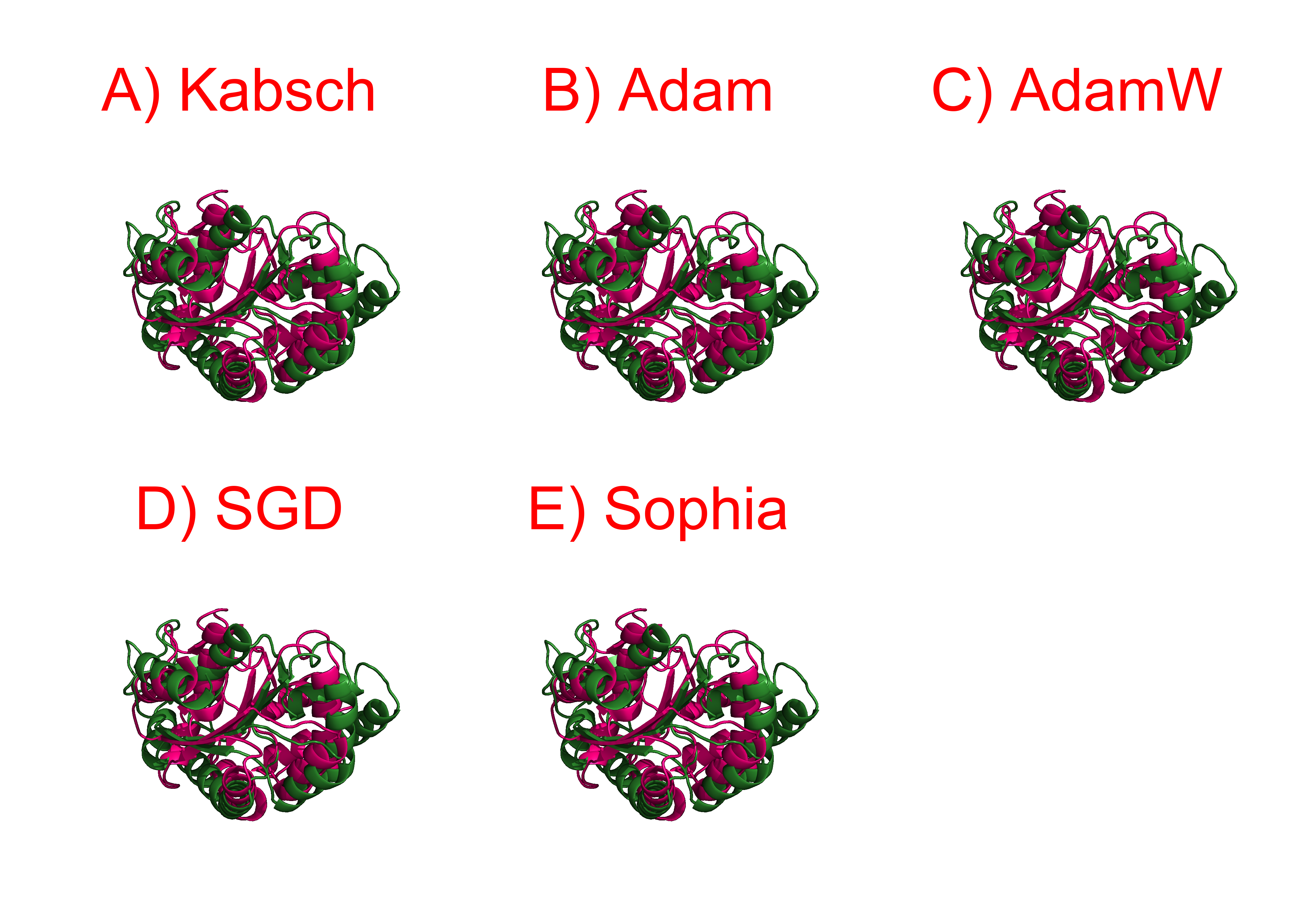}
\caption{Visual comparison of the structural alignments produced by Kabsch and the four gradient-based optimizers. The reference structure (1AKE, pink) and the aligned mobile structure (4AKE, green) are shown. All methods produce visually identical superpositions, validating the convergence of the gradient-based methods.}
\label{fig:comparison}
\end{figure}

The visual representation of these alignments (Figure \ref{fig:comparison}) confirms the numerical results. All methods produce superpositions that are visually indistinguishable from one another, perfectly overlaying the mobile structure (4AKE) onto the reference structure (1AKE).

\section{Discussion}
The key finding of this work is that a differentiable structural alignment framework using Lie algebra can robustly and precisely replicate the analytical solution of the Kabsch algorithm. This is a crucial result, as it validates that the Lie algebra parameterization of SE(3) creates a well-behaved optimization landscape for RMSD that modern optimizers can successfully navigate to find the global minimum.

\subsection{Comparison of Methods}
The convergence of all four optimizers is expected, as the RMSD minimization problem is convex. For such a problem, gradient descent with an appropriate learning rate is guaranteed to find the global optimum. The minor differences in their final RMSD values are numerically insignificant and stem from the stochastic nature of the optimizers and floating-point arithmetic.

The most striking difference is in computational time. The Kabsch algorithm, being a direct analytical solution, is three orders of magnitude faster than the iterative gradient-based methods. This is because it solves the problem in a single pass of matrix operations, whereas optimizers must perform hundreds of iterative forward and backward passes (gradient calculations). This highlights that for pure RMSD minimization, the analytical approach remains superior in performance.

However, the purpose of Lie-RMSD is not to outperform Kabsch in its own domain. The power of our framework lies in its flexibility and extensibility.

\subsection{Implications for Complex Scoring Functions}
Having validated the framework's accuracy on a classic problem, we can now confidently apply it where analytical methods fail. The simple, differentiable RMSD loss function can be replaced with more complex, potentially non-differentiable, or bespoke scoring functions. For instance, one could design a loss function that combines RMSD with terms for:
\begin{itemize}
    \item \textbf{Biophysical properties:} Penalizing clashes, rewarding favorable electrostatic or hydrophobic interactions.
    \item \textbf{Topological similarity:} Incorporating metrics like the TM-score or GDT-TS, which are more robust to outliers and flexible regions.
    \item \textbf{Data-driven scores:} Using potentials derived from deep learning models trained to recognize native-like protein interfaces.
\end{itemize}
For these complex objectives, no analytical optimizers exist. Our framework provides a direct, robust, and validated path to finding meaningful structural alignments under arbitrary, user-defined criteria. This work therefore bridges the gap between classical analytical methods and modern, flexible gradient-based optimization for a wide range of problems in structural biology.

\section{Conclusion}
We have introduced and validated Lie-RMSD, a fully differentiable framework for protein structure alignment based on a Lie algebra representation of rigid-body motion. Our experiments demonstrate that this approach achieves precision effectively identical to the seminal Kabsch algorithm when minimizing RMSD, confirming its correctness and robustness. While not as fast as the analytical method for this specific task, the Lie-RMSD framework provides a solid and validated foundation for future work on optimizing complex, non-analytical scoring functions that are of growing importance in structural bioinformatics.

\section*{Acknowledgements}
The authors would like to acknowledge the assistance of Google's Gemini model in the preparation of this manuscript. Gemini provided support in writing and testing the Python code, as well as in refining the language and structure of the paper. All code, experimental results, and the final manuscript were reviewed and verified by the human authors, who take full responsibility for the content of this work.


\begin{thebibliography}{9}
\bibitem{ref1}
Koehl, P. (2010). Protein structure similarities. \textit{Current Opinion in Structural Biology}, 20(3), 392-397.

\bibitem{kabsch1976}
Kabsch, W. (1976). A solution for the best rotation to relate two sets of vectors. \textit{Acta Crystallographica Section A}, 32(5), 922-923.

\bibitem{kabsch1978}
Kabsch, W. (1978). A discussion of the solution for the best rotation to relate two sets of vectors. \textit{Acta Crystallographica Section A}, 34(5), 827-828.

\bibitem{ref2}
Betancourt, M. R., \& Skolnick, J. (2001). Universal similarity measure for comparing protein structures. \textit{Biopolymers}, 59(5), 305-309.

\bibitem{tmalign}
Zhang, Y., \& Skolnick, J. (2005). TM-align: a protein structure alignment algorithm based on the TM-score. \textit{Nucleic acids research}, 33(7), 2302-2309.

\end{thebibliography}
\end{document}